\documentclass[aps,prl,twocolumn,groupedaddress,superscriptaddress,showpacs]{revtex4}

\usepackage{graphicx}
\usepackage{mathrsfs}

\def\vJ{{\bf J}}
\def\vS{{\bf S}}
\def\vl{{\bf l}}
\def\vr{{\bf r}}
\def\vs{{\bf s}}
\def\vE{{\bf E}}

\begin{document} 

\title{All-electrical control of single ion spins in a semiconductor} 

\author{Jian-Ming Tang}
\affiliation{Department of Physics and Astronomy, University of Iowa, Iowa City, IA 52242-1479}
\author{Jeremy Levy}
\affiliation{Department of Physics and Astronomy, University of Pittsburgh, Pittsburgh, Pennsylvania 15260, USA}
\author{Michael E. Flatt\'e}
\affiliation{Department of Physics and Astronomy, University of Iowa, Iowa City, IA 52242-1479}

\begin{abstract}

  We propose a method for all-electrical manipulation of single ion
  spins substituted into a semiconductor. Mn ions with a bound hole in
  GaAs form a natural example. Direct electrical manipulation of the
  ion spin is possible because electric fields manipulate the orbital
  wave function of the hole, and through the spin-orbit coupling the
  spin is reoriented as well. Coupling ion spins can be achieved using
  gates to control the size of the hole wave function. Coherent
  manipulation of ionic spins may find applications in high density
  storage and in scalable coherent or quantum information processing.

\end{abstract}


\maketitle 

Observing magnetic resonance between different spin states of nuclei
or electrons is a technique widely used in many imaging and
spectroscopic applications. Sensitivity sufficient to measure the
fluctuation of a single spin has been demonstrated using magnetic
resonance force microscopy~\cite{Rugar2004}, noise
spectroscopy~\cite{Xiao2004}, optical
spectroscopy~\cite{Gruber1997,Jelezko2002}, scanning tunneling
microscopy (STM)~\cite{Manassen1989,Durkan2002,Heinrich2004}, and
quantum point contact conductivity~\cite{Elzerman2004}. The potential
also exists to determine the spin
state~\cite{Tang2005,Jelezko2002,Xiao2004,Elzerman2004}. Controlling a
single spin, in addition to monitoring it, is highly desirable for
building future spin-based devices, is essential for quantum
computation, and should permit the direct exploration of fundamental
aspects of quantum dynamics in a solid state
environment~\cite{Prinz1998,Wolf2001,Ziese2001,Awschalom2002b}. Proposed
schemes to control a single spin in a solid state environment rely
either on magnetic resonance~\cite{Loss1998,Kane1998} or optical
manipulation~\cite{Imamoglu1999,Pryor2002,Quinteiro2005}, and there has
been progress towards replacing magnetic control fields with the
spin-orbit interaction~\cite{Kato2003,Kato2004} or the exchange
interaction~\cite{DiVincenzo2000,Petta2005}.

Ionic spin states in solids have several attractive characteristics
for fundamental studies of spin dynamics and for spin-based
devices. Every ion embedded in a solid is identical to every other
such ion. Thus an ionic spin
system can be as uniform as a nuclear spin system, but also can permit
spin manipulation on short time scales as in a quantum dot spin
system. Controlling ionic single spins
without any magnetic fields, using techniques in which electric fields
play the typical role of magnetic fields, may therefore provide a path
to high-density scalable spin-based electronics.  For example, the
control of ionic spin states can be used to produce highly
spin-selective and spin-dependent tunneling currents in nanoscale
electrical devices, or to realize quantum computation. 
Manipulation of individual spins that are constituents of interacting
spin clusters also opens up the capability to explore the fundamental
dynamics of frustrated spin systems and other correlated spin systems.

Here we propose an all-electrical scheme for ionic spin manipulation
in which the role of magnetic fields in traditional electron spin
resonance (ESR) is replaced by electric fields.  In conventional ESR
the energy splitting between different spin states, and the couplings
between them, are controlled by magnetic fields because an electric
field does not directly couple to the electron's spin. In a
semiconductor crystal with tetrahedral symmetry and spin-orbit
interaction (such as GaAs) a $J=1$ ion spin (such as that of Mn in GaAs) will
be triply degenerate, however the energy splittings and the couplings
between these states depend {\em linearly} on the {\em electric field}
strength, allowing rapid all-electrical control. Thus all operations
performed with magnetic fields in traditional ESR, can be performed
with electrical techniques.

\begin{figure}[b]
\includegraphics[width=\columnwidth]{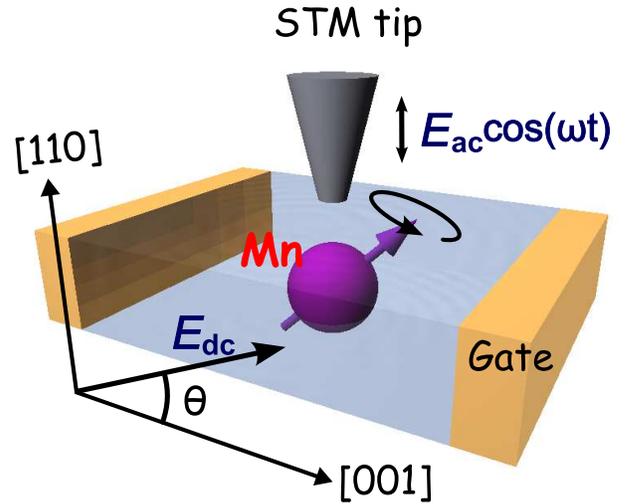}
\caption{(color online) Proposed configuration for the electric resonances of a
single Mn dopant in GaAs. A dc electric field $\vE_{dc}$ is applied
via the electrical gates and the STM tip. The resonance is driven by
an additional small ac field. }
\label{fig:STM}
\end{figure}

A specific proposed setup for manipulating a single ion spin is shown
in Fig.~\ref{fig:STM}. Tip-induced placement of single Mn ions
substituted for Ga in a GaAs sample has been demonstrated
experimentally~\cite{Kitchen2005}.  Two gates are configured to apply
an electric field along the $[001]$ axis. The STM tip serves as the
third gate for spin manipulation, and as a contact for initialization
and detection. Taking advantage of the $(110)$ natural cleavage plane
(which lacks surface states), the applied electric field is confined
in the $(1\bar{1}0)$ plane and the orientation is specified by the
angle $\theta$ from the $[001]$ axis.

An isolated Mn atom has a half-filled $3d$ shell and the spins of all
five $3d$ electrons are aligned (Hund's rule) to form a $S=5/2$ ground
state. In GaAs a hole in the valence
band compensates for the differing valences of Mn and Ga.
We describe the core spin-valence hole dynamics with the following
effective spin Hamiltonian:
\begin{eqnarray}
{\mathscr H}_{\rm spin} & = & \alpha\vS\cdot\vs + \beta\vl\cdot\vs \;,\label{first-spin-hamiltonian}
\end{eqnarray}
where $\vl$ and $\vs$ are the orbital angular momentum ($l=1$) and the
spin of the bound hole respectively. Our tight-binding
calculations~\cite{Tang2004} estimate the exchange coupling $\alpha$
and the spin-orbit coupling $\beta$ to be about $300$~meV and
$-80$~meV respectively. The exchange interaction binds the valence
hole with spin antiparallel to the Mn core spin with a binding energy
of $113$~meV~\cite{Lee1964}.  The spin-orbit interaction in GaAs
configures the orbital angular momentum of the hole parallel to its
spin. The total angular momentum of the (Mn core + hole) complex is
$\vJ = \vS + \vl + \vs$, and the ground state of this complex has
$J=1$~\cite{Schneider1987} (both $\vl$ and $\vs$ are antiparallel to
$\vS$),  confirmed via ESR~\cite{Schneider1987}. Our
proposals for spin control involve energy scales smaller than
$\alpha$ or $\beta$, so only the lowest
energy multiplet with $J=1$ is of interest here.

The degeneracy of the $J=1$ Mn ion can be substantially split by
external electric fields, and the eigenstates depend strongly on the
electric field direction. This will be the source both of state
splitting (analogous to the static magnetic field in traditional ESR)
and state coupling (analogous to the oscillating perpendicular
magnetic field in traditional ESR). We find the following
electric-field-dependent Hamiltonian:
\begin{eqnarray}
{\mathscr H}_I(\vE) & = & \gamma \left[
E_x(J_yJ_z+J_zJ_y)+{\rm c.p.}\right] \;, \label{eq:H}
\end{eqnarray}
where $\vE$ is an electric field, c.p. stands for cyclic
permutation, and $\{x,y,z\}$ stand for the 3 major axes of the cubic
crystal. Note that this Hamiltonian does not break time-reversal
symmetry, for the angular momentum operators $\vJ$ always appear in
pairs. We calculate, using the probability densities of the hole state
found in our tight-binding calculations and first-order perturbation
theory, $\gamma = 6.4\times 10^{-30}$~Cm, corresponding to $\gamma E
=160$~$\mu$eV for $E=40$~kV/cm. This exceptionally large splitting is
equivalent to that generated by applying a $1$ Tesla magnetic field
using the measured $g$-factor~\cite{Schneider1987}, $2.77$.  The
linear dependence on electric field, critical to producing a large
splitting, originates from the lack of inversion symmetry of the
substituted ion in a tetrahedral host. The energy splittings from an
electric field applied to bound states at inversion-symmetric sites in
crystals, or electrons bound in atoms or ions in vacuum, would depend
quadratically on the electric field and would be correspondingly much
smaller.  The other essential element causing this large splitting is
the large ($\sim 10$\AA) Bohr radius of the bound valence
hole~\cite{Tang2004,Yakunin2004}. Recent progress in theory and
scanning tunneling microscopy of Mn dopants in III-V semiconductors
has confirmed the large spatial extent of the bound hole
wavefunction~\cite{Arseev2003,Tang2004,Yakunin2004,Kitchen2005}. Thus
the response of the Mn wavefunction to electric fields is substantial
compared to other ion levels associated with transition-metal
(magnetic) dopants.

\begin{figure}[b]
\centerline{\includegraphics[width=\columnwidth]{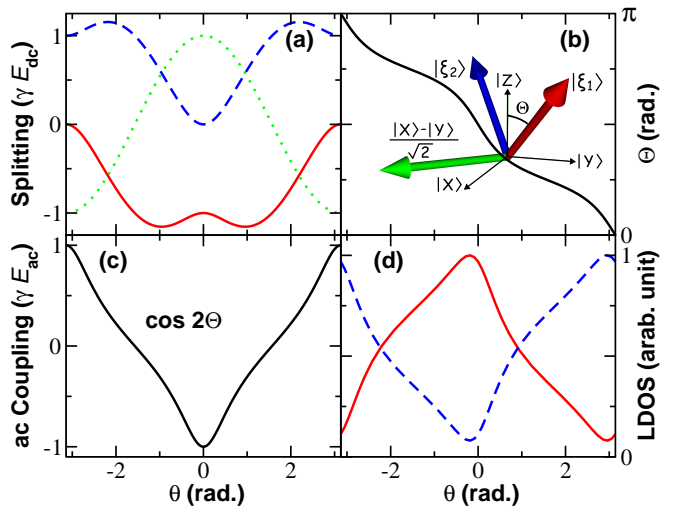}}
\caption{(color online) The ionic spin system as a function of the dc
field orientation. (a) The energies of the $J=1$ states $\xi_1$ (solid), $\xi_2$ (dashed), and $\xi_3$ (dotted).  (b)
The corresponding eigenvectors parametrized by the angle $\Theta$. (c) The coupling between $|\xi_1\rangle$ and $|\xi_2\rangle$ due to the ac field. (d) The scaled LDOS of the two
possible final states $|\xi_1\rangle$ (solid) and $|\xi_2\rangle$ (dashed) probed four monoatomic layers directly above the Mn
dopant.  }
\label{fig:dc_xxz}
\end{figure}

In the basis $|X\rangle$, $|Y\rangle$ and $|Z\rangle$, defined by
$J_\alpha|\alpha\rangle = 0$, the Hamiltonian can be written as
\begin{eqnarray}
{\mathscr H}_I(\vE) & = & -\gamma E\left(\begin{array}{ccc}
0 & \hat E_z & \hat E_y \\
\hat E_z & 0 & \hat E_x \\
\hat E_y & \hat E_x & 0
\end{array}\right) \;.
\label{eq:Hxyz}
\end{eqnarray}
The energy eigenvalues in units of $\gamma E$ are the roots of the
characteristic polynomial,
\begin{eqnarray}
x^3 - x + 2\eta & = & 0 \;,
\end{eqnarray} 
where $\eta = \hat E_x \hat E_y \hat E_z$. A static electric field
$\vE_{dc}$ splits all three eigenstates in energy except when the
field is in the $[111]$ direction (or equivalent), for which two of
the eigenstates remain degenerate.

The energies of the three states are
$\xi_1=(-\cos\theta-\sqrt{4-3\cos^2\theta})/2$,
$\xi_2=(-\cos\theta+\sqrt{4-3\cos^2\theta})/2$, and
$\xi_3=\cos\theta$, shown by the solid, dashed, and dotted curves
respectively in Fig.~\ref{fig:dc_xxz}(a). The eigenstate
$|\xi_3\rangle=(|X\rangle-|Y\rangle)/\sqrt{2}$ is independent of $\theta$.
The independence of $|\xi_3\rangle$ from $\vE$ (in this geometry) motivates
us to define a pseudospin $1/2$ constructed from the other two states,
$|\xi_1\rangle$ and $|\xi_2\rangle$. These eigenstates can be written
as
$|\xi_1\rangle=(\sin\Theta/\sqrt{2},\sin\Theta/\sqrt{2},\cos\Theta)$ and
$|\xi_2\rangle=(-\cos\Theta/\sqrt{2},-\cos\Theta/\sqrt{2},\sin\Theta)$, where
$\Theta$ is the angle of between $|\xi_1\rangle$ and the $|Z\rangle$
basis (Fig.~\ref{fig:dc_xxz}(b)). Note that all the eigenvectors are
real because of time-reversal symmetry.

\begin{figure}[b]
\includegraphics[width=\columnwidth]{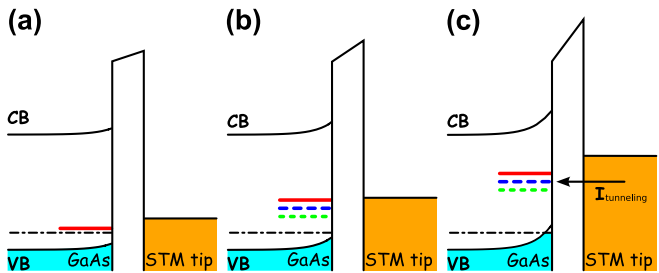}
\caption{(color online) Schematics of controlling the spin states via
local band bending. The dot-dashed lines show the
chemical potential. Shaded regions are filled states. CB and VB label the conduction and valence bands of the semiconductor. (a) Initialization.  For this voltage occupation of the $|\xi_1\rangle$ state dominates. (b) Manipulation: Bring all the states into
the gap, but control the bias voltage below the threshold where the
current starts to tunnel through these states. The oscillating field
($E_{ac}$) drives transitions between the $|\xi_1\rangle$ and the
$|\xi_2\rangle$ states. (c) Detection: Bring the final state further
into the gap, so that electrons can tunnel from the tip into the
acceptor state. The final state is identified according to the
amplitude of the tunneling current (Fig.~\ref{fig:dc_xxz}(d)). }
\label{fig:init}
\end{figure}

Preparation of the initial pseudospin state is achieved by applying an
electric field to split the state energies, and allowing the hole to
relax into the ground state. The electric field from the STM tip
locally bends the bands of the semiconductor and permits ionization of
the bound hole; this has been demonstrated for Mn in
GaAs~\cite{Yakunin2004,Yakunin2004b}.  Rapid initialization of a high
purity pseudospin state can be achieved by using the local band
bending effect to move the two higher-energy levels ($\xi_2$, $\xi_3$)
to the position shown in Fig.~\ref{fig:init}(a), so a hole in those
states would ionize and be replaced by a hole in the lowest energy
state ($|\xi_1\rangle$). At a temperature of 0.5~K and a dc field of
100~kV/cm, the occupation of the next highest state ($|\xi_2\rangle$)
would be less than $10^{-4}$. We have chosen $\vE_{dc}$ such that
$|\theta|<(\pi-\tan^{-1}\sqrt{2})$, so that $|\xi_1\rangle$ (not
$|\xi_3\rangle$) is the ground state (see Fig.~\ref{fig:dc_xxz}(a)).
Band bending also changes the effective radius of the bound hole wave
function; gate voltages applied at the surface could thus control the
coupling of two bound hole states in an analogous way to approaches in
Refs.~\cite{Loss1998,Kane1998} for quantum dots and donor states.

In order to manipulate the initialized spins the tip-sample bias
should be increased adiabatically (slower than $\hbar/(\gamma
E_{dc})$) to bring all three levels into the semiconductor's energy
gap (see Fig.~\ref{fig:init}(b)). This shift with bias is described
for Mn in $p$-doped GaAs in Ref.~\cite{Yakunin2004}. The bias voltage
has to be maintained below the critical value at which electrons start
to tunnel directly through these levels, so that the transitions
between these states remain coherent. Spin resonance can now be driven
by applying a small oscillating electric field $\vE_{ac}(t)$ to the
static field $\vE_{dc}$. The Hamiltonian
\begin{eqnarray}
{\mathscr H}_{\rm ESR} & = & {\mathscr H}_I(\vE_{dc}) + {\mathscr H}_I[\vE_{ac}(t)] \;.
\end{eqnarray}
To have a well-defined pseudospin $1/2$, constructed out of
$|\xi_1\rangle$ and $|\xi_2\rangle$, the coupling of these two states
to $|\xi_3\rangle$ must vanish. For the schematic in
Fig.~\ref{fig:STM} the oscillating field can be applied either along
the $[110]$ direction through the STM tip or along the $[001]$
direction through the gates. Both choices leave $|\xi_3\rangle$
unaffected and only couple $|\xi_1\rangle$ and $|\xi_2\rangle$ to each
other.  To see how the states are coupled by the ac field, we write
out ${\mathscr H}_I[\vE_{ac}(t)]$ using the eigenstates of ${\mathscr
H}_I(\vE_{dc})$ as bases. We assume that the ac field $\vE_{ac}(t)$ is
along the $[110]$ direction.
\begin{eqnarray}
\mathscr{H}_I[\vE_{ac}(t)] & = & \gamma E_{ac}\cos(\omega t)\left(\begin{array}{ccc}
-\sin2\Theta & \cos2\Theta & 0 \\
\cos2\Theta & \sin2\Theta & 0 \\
0 & 0 & 0
\end{array}\right) \;.
\end{eqnarray}
The off-diagonal term $\cos2\Theta$, plotted in
Fig.~\ref{fig:dc_xxz}(c), shows how the coupling between the two
coupled states changes with the field orientation. The coupling is
maximized when the static field is completely along the $[001]$
direction ($\theta=0$). Then the static and oscillating electric
fields are perpendicular to each other, just as the static and
oscillating magnetic fields are perpendicular to each other in
traditional ESR. In the limit $E_{ac} \ll E_{dc}$, the diagonal term
can be neglected and our configuration works just like conventional
ESR. The Rabi frequency obtained from the standard Rabi formula is
\begin{eqnarray}
\hbar\Omega & = & \frac{1}{2}\sqrt{(\gamma E_{ac}\cos2\Theta)^2+(\hbar\omega-\gamma E_{dc}\sqrt{4-\cos^2\theta})^2} \;. \nonumber\\
\end{eqnarray} 
For $E_{ac} = E_{dc}/4 = 25$~kV/cm, and $\Theta=\pi/2$, $\Omega/2\pi =
12$~GHz, corresponding to a Rabi time of 80~ps. Ensemble spin
coherence times $T_2^*$ measured by traditional ESR in GaMnAs exceed
$0.5$~ns (several times the estimated Rabi time), and appear due to
the inhomogeneous environments of Mn ions~\cite{Schneider1987}; the
$T_2$'s of individual spins are expected to be considerably
longer. Hyperfine interactions, which significantly affect conduction
electron spin coherence, are expected to be weak for Mn ions as the
overlap of the valence $p$ orbitals with the nucleus is small.

High-fidelity determination of the orientation of the pseudospin can
be achieved by measuring the total tunneling current through the final
state with the STM~\cite{Tang2005}.  When the tip-sample voltage is
increased, and the semiconductor bands bend further (see
Fig.~\ref{fig:init}(c)), current starts to tunnel through the bound
hole wavefunction state~\cite{Yakunin2004,Yakunin2004b} and the
tunneling current is proportional to the probability density of the
state at the STM tip location. The spatial structure of these $J=1$
states is highly
anisotropic~\cite{Tang2004,Yakunin2004,Yakunin2004b}. We calculate the
two eigenstates to have the following spatial structure,
\begin{eqnarray}
\langle\vr|\xi_i\rangle & = & c^i_{XY}\langle\vr|X+Y\rangle + c^i_Z\langle\vr|Z\rangle \;,
\end{eqnarray}
where $c^1_{XY}=\sin\Theta$, $c^2_{XY}=-\cos\Theta$, and $|X+Y\rangle=(|X\rangle+|Y\rangle)/\sqrt{2}$. If the STM tip
is positioned directly over the Mn dopant, it probes the nodal plane,
$\langle\vr|Z\rangle = 0$. In this particular case, the local density of states (LDOS) is simply
proportional to the projection on to the $\langle\vr|X+Y\rangle$
state, which is $|\sin\Theta|^2$ for $|\xi_1\rangle$, and
$|\cos\Theta|^2$ for $|\xi_2\rangle$.  Thus for a static electric
field along $\theta=0$ ($\Theta=\pi/2$) and the pseudospin in the
$|\xi_2\rangle$ state no current will be detected, but for the
pseudospin in the $|\xi_1\rangle$ state there will be current
detected. The difference in current for the $|\xi_1\rangle$ and
$|\xi_2\rangle$ states is shown in Fig.~\ref{fig:dc_xxz}(d). For this
position about 10\% of the LDOS is not spin dependent, which reduces
maximum visibility to $90$\% (at, e.g., $\theta\approx0$). Spatial
averaging of the LDOS over a typical experimental $2$\AA\ changes the
visibility by only a few percent. The asymmetric angular dependence is
due to the lack of inversion symmetry of the substituted ion in a
tetrahedral host. Current measurement timescales can be very fast as
STM experiments performed at 50~GHz have
demonstrated~\cite{Steeves1998}.  We also assume that the tunneling
current is small so that spin flip does not occur during the
measurement.

Controllable coupling of two spins permits use of these Mn ions for
quantum information processing. Estimates of the overlap of holes
bound to two separated Mn ions~\cite{Tang2004} indicate $\sim 100$~meV
splittings of Mn pair states for ions separated by 12~\AA\ along the
$(1\bar{1}0)$ direction. The overlap falls off for larger
separations according to the $\sim 13$~\AA\ wave function radius of
the bound hole~\cite{Tang2005}, so it would be $\sim 0.1$~meV for two ions 10~nm
apart. This overlap could be reduced, increased, or eliminated with a
gate between the two ions~\cite{Loss1998,Kane1998}. By using single-Mn
manipulations to put single-ion quantum information in the proper pair
of single-Mn states, the Mn pair state splitting can be used to
perform CNOT operations in an analogous way to how the singlet-triplet
splitting is used for a CNOT with spin$-1/2$ qubits.

In conclusion, we have presented a concrete proposal for electrically
initializing, manipulating, and detecting single pseudospin states of
a magnetic dopant in a semiconductor. All-electrical spin manipulation
should be possible for other impurities in tetrahedral semiconductors
characterized by $J>1/2$ ground state spins (e.g. most transition
metal ions in most tetrahedral semiconductors, or the nitrogen-vacancy
center in diamond). In a future scalable architecture the STM tip
would be replaced by a gate-controlled contact. The controlled
resistance of that contact would permit alternation between the gate
configuration for manipulation and the contact configuration for
initialization and detection, all without moving parts. The $[001]$
static electric field, here assumed to be implemented with gates, may
also be replaced by an internal electric field from a doping gradient
(such as in a $p-n$ junction), or even a static strain field. The Mn
ions could be controllably placed within the surface relative to the
contacts using current pulses from an STM tip as described in
Ref.~\cite{Kitchen2005}.

This work is supported by ARO MURI DAAD19-01-1-0541 and DARPA QuIST
DAAD-19-01-1-0650.

%\bibliography{semiconductor}

\end{document}